\documentclass[pra,twocolumn,showpacs,amsmath,amssymb,superscriptaddress]{revtex4}

\usepackage{amssymb,amsfonts,amstext}
\usepackage{amsmath}	
\usepackage{latexsym}
\usepackage{color}
\usepackage{multirow}
\usepackage{algorithm}
\usepackage{algorithmic}

\newcommand{\ket}[1]{| #1 \rangle}

\newcommand{\be}{\begin{equation}}
\newcommand{\ee}{\end{equation}}

\def\QED{\mbox{\rule[0pt]{1.5ex}{1.5ex}}}
\def\endproof{\hspace*{\fill}~\QED\par\endtrivlist\unskip}

\def\Label#1{\label{#1}\ [\ \text{\protect{#1}}\ ]\ }
\def\Label{\label}

\begin{document}
\title{Arbitrable Blind Quantum Computation}
\author{Go Sato}
\affiliation{Division of Mathematics Electronics and Informatics,
Graduate School of Science and Engineering, Saitama University, 255
Shimo-Okubo, Sakura-ku, Saitama 338-8570, Japan}
\author{Takeshi Koshiba}
\affiliation{Faculty of Education and Integrated Arts and Sciences,
Waseda University,
1-6-1 Nishiwaseda, Shinjuku-ku, Tokyo 169-8050, Japan}
\author{Tomoyuki Morimae}
\affiliation{Department of Computer Science, Gunma University, 1-5-1 Tenjin-cho, Kiryu-shi, Gunma 376-0052, Japan}


\begin{abstract}
Blind quantum computation is a two-party protocol which involves
a server Bob who has rich quantum computational resource
and provides quantum computation service and a client Alice who
wants to delegate her quantum computation to Bob without revealing her quantum
algorithms and her input to (resp., output from) the algorithms.
Since Bob may be truant and pretend to execute some computation, Alice
wants to verify Bob's computation. Verifiable blind quantum computation
enables Alice to check whether Bob is cheating or not. If Bob is
cheating and claims his innocence, Alice can refute the denial of
Bob's cheating but she cannot persuade any others that Bob is cheating.
In this paper, we incorporate arbitrators 
as the third party into blind quantum computation 
to resolve the above problem and give an
arbitrable blind quantum computation scheme.
\end{abstract}

\maketitle

\section{Introduction}
Secure computation involves several parties who want to 
evaluate functions over their private inputs 
without compromising the privacy.
Since the notion of secure computation was invented by Yao,
there have been many proposals to implement secure computation.
The first scheme by Yao \cite{Yao86} was realized as a combination
of encrypted circuits (garbled circuits) to evaluate a function 
and oblivious transfer protocols and its security relied on
an unproven computational complexity theoretic assumption.
Unfortunately, unconditionally secure computation for general functionalities
is impossible in the classical setting \cite{Cleve86} and
even in the quantum setting \cite{Lo97}.
Thus, we have to consider special cases where 
unconditionally secure computation can be realized.
As a special case of secure computation, a protocol (BFK09 protocol) for 
unconditionally secure delegated computation was shown by Broadbent, 
Fitzsimons and Kashefi \cite{BFK09}. Secure delegated computation
in the quantum setting (a.k.a. blind quantum computation (BQC)) is
a two-party protocol between a client (Alice) and a server (Bob):
Bob is capable of carrying out any quantum computation and Alice
wants to delegate her computational tasks to Bob without revealing
her secret, that is, her algorithms and their inputs/outputs.
After Broadbent et al's seminal work, BQC protocols in different settings
and model have been proposed \cite{MF13,SKM13,MF12,Morimae12,DKL12,GMMR13,MPF13,DFPR14,MDK15,MF13a}.

While BQC protocols ensure Alice's privacy, Alice may not detect
Bob's cheating behaviors if Bob neglects Alice's request and
pretends to execute her algorithm. To overcome this problem, some BQC
protocols support the {\em verifiability} which enables to detect Bob's
dishonest behavior if Bob is cheating. The Fitzsimons-Kashefi protocol
(FK12 protocol) \cite{FK12} is a verifiable variant of the BFK09 one
and the Hayashi-Morimae protocol (HM15 protocol) \cite{HM15} is 
a verifiable variant of the Morimae-Fujii BQC one (MF13 protocol) \cite{MF13}.
If Alice can verify Bob's execution of Alice's algorithm, there still exists 
a problem. Even if Alice finds Bob's cheating behavior, Bob may insist that
he is innocent and that Alice rather tries to ensnare him. 
For example, Alice pays a service fee to Bob if Bob honestly provides
the delegated computation service. On the other hand, Alice does not
want to pay anything if Bob is cheating. The cheating Bob may charge Alice
for his cheating service.
Any person outside the protocol (e.g. the FK12 protocol
or the HM15 protocol) cannot decide which one is honest. 

The notion of verifiability given in \cite{FK12} is what we should call it {\em private
verifiability}. We have observed that the private verifiability does not 
resolve the above dispute.
To resolve this dispute problem, we rather need a mechanism 
where any public outsider or the specific third-party can
verify if Bob honestly provides a delegated computation service.
In \cite{Honda16}, Honda proposes 
the notion of {\em public verifiability} for BQC
and provides a publicly verifiable BQC protocol by using a classical
computational cryptography, which relies on an unproven 
computational assumption. In this paper, we take a different
approach from \cite{Honda16} and provide a protocol which does not
rely on computational assumptions. To do this, we 
incorporate arbitrators as the third party into 
the HM15 protocol and give an unconditionally-secure
{\em arbitrable} blind quantum computation protocol.

For BQC, there is another approach by using homomorphic encryptions.
In the classical setting, Gentry devised a fully homomorphic 
encryption \cite{Gentry09}. Even in the quantum settings,
some possibilities of homomorphic encryptions have been 
discussed \cite{BJ15,DSS16}.

\section{Preliminaries}
As mentioned, we will give an arbitrable blind quantum computation protocol.
First, we briefly review the MF13 protocol on which we base our protocol. 
In our protocol, we need a procedure (i.e., honesty test) to decide 
if a given state is actually a graph state. 
For such an honesty test, we may use 
the Morimae-Nagaj-Schuch test (MNS16 test) 
\cite{MNS16} or tests in \cite{MTH17,TM17,HM15}
and the security can be proved by using quantum de Finetti 
Theorem under measurements that are implementable by
fully-one-way local operations and classical communication (LOCC)
\cite{LS15}.

\subsection{MF13 and HM15 protocols}
A client Alice has only a measurement device and a server Bob prepares
a resource state of measurement-based quantum computation. 
For the MH13 protocol, any resource state (e.g., 
two-dimensional cluster state or three-dimensional cluster state
for topological quantum computation) can be used. 
The MH13 protocol is as follows: (1) Bob prepares a universal 
resource state; (2) Bob sends a particle of the resource state to Alice
via the quantum channel; (3) Alice measures the particle 
with respect to the basis which is determined by her algorithm.
They repeat (2) and (3) until the computation halts.

The HM15 protocol is a privately-verifiable version of the MF13 protocol.
In the HM15 protocol, Bob first prepares $k+1$ copies of the
resource state (represented as a bipartite graph). $k$ 
out of $k+1$ copies are used for stabilizer tests for the verifiability.
The remaining one copy is used for the computation where
Alice's algorithm is performed via the MF13 protocol.

\subsection{Honesty Test}
Morimae, Nagaj and Schuch \cite{MNS16} gave a verification
procedure (MNS16 test) that checks whether a given quantum 
state $|\psi\rangle$ is close to a graph state $|G\rangle$. The probability
the MNS16 test passes is described as
$(1 + \langle \psi | G \rangle)/2$. 
While the honesty test in \cite{HM15} is for bipartite-graph states,
the MNS16 test and also tests in \cite{MTH17,TM17} are
for any graph states. 
In the protocol description we will show, we mention the MNS16 test
for the honesty test.
Alternatively, we may use other tests 
in \cite{MTH17,TM17,HM15}.

\section{Protocol}\Label{S4}
As in the standard BQC protocols, we assume that Alice, a client, 
would like to securely delegate her computation to Bob, a server.
We assume that Bob can prepare a universal $n$-qubit graph state $|G\rangle$.
Besides Alice and Bob, we assume that there exists a trusted
third-party Charlie who acts as an arbitrator. In this paper,
we assume that Charlie always obeys the protocol and stands neutral.
Our protocol is given in Protocol 1.

For the protocol, we assume that
$k\ge 4n^2 -1$ and $m\ge (2\ln 2) k^nn^5$. 
It is not essential that Charlie has quantum memory in the protocol.
If Charlie does not have quantum memory, he can take the option in STEP 3.
Moreover, if Alice considers that her private verification suffices,
Charlie does not have to be involved in the protocol. In that case,
Bob directly sends $k+m+1$ graph states to Alice in STEP 1;
Alice applies a random permutation to $k+m+1$ graph states and discards $m$
graph states in STEP 2; and STEP 3 and STEP 6 can be omitted.

\begin{Protocol}                  
\caption{Arbitrable BQC protocol}\label{P:abqc}
\begin{algorithmic}
\STEPONE
Bob generates $\ket{G}^{\otimes 2k+m+1}$
and sends them to Charlie, where
$\ket{G}$ is an $n$-qubit graph state.

\STEPTWO
Charlie applies a random permutation to $2k+m+1$ graph states
and discards $m$ graph states.

\STEPTHREE
Charlie keeps $k$ graph states in his memory and
sends the remaining $k+1$ graph states $\rho$ to Alice.
(Optionally, Charlie may execute $k$ MNS16 tests as in {STEP 6}.
If the $k$ tests are not passed, Charlie judges that Bob is cheating.)

\STEPFOUR
Alice receives a graph states $\rho$
and applies the MNS16 tests to randomly chosen $k$ graph states
from $k+1$ graph states $\rho$. Let $\rho_{comp}$ be the
remaining one graph state.
Alice executes the algorithm on $\rho_{comp}$
by measuring the particle with respect to the basis which is
determined by the description of Alice's algorithm. 

\STEPFIVE
If those $k$ tests pass,
then Alice accepts the computation.
Otherwise, Alice rejects the computation.
Charlie does not anything if Alice accepts.

\STEPSIX
If Alice rejects, then Charlie
executes $k$ MNS16 tests by using $k$ graph states stored in his memory. 
If those $k$ tests are passed, Charlie
judges that Alice is cheating. Otherwise, Charlie
judges that Bob is cheating.
\end{algorithmic}
\end{Protocol}

Then, Protocol 1 satisfies the following two properties.\\
({\it Completeness})\\
If Bob sends $|G\rangle^{\otimes 2k+m+1}$ to Alice (via Charlie), then
Alice passes the test with probability 1.\\
({\it Soundness})\\
If Alice passes the test, $\rho_{comp}$ satisfies
$\langle G | \rho_{comp} | G\rangle \ge 1-\frac{1}{n}$ with
probability $1 - \frac{1}{n}$ at least.

The completeness just comes from the construction of Protocol 1.
The soundness can be similarly discussed as in \cite{MTH17}.
For self-containment, we review a proof in Appendix.

On Eq.(\ref{eq}) in Appendix, we suppose that $T$ is a POVM for Alice
in STEP 5. On the other hand, we can provide another interpretation
for Eq.(\ref{eq}) and consider that $T$ is a POVM for Charlie
in STEP 6. Even in this interpretation, we have a similar consequence. 
That is,
if Charlie passes the test, $\rho_{comp}$ satisfies $\langle G | \rho_{comp}
| G\rangle\ge 1-\frac{1}{n}$ with probability $1-\frac{1}{n}$ at least.
In other words, if Alice accepts the computation, then Charlie can
endorse her acceptance. Thus, we can say that Charlie works as an arbitrator.

\section*{Acknowledgments}
TK is supported in part by JSPS Grant-in-Aids for Scientific
Research (A) 16H01705 and for Scientific Research (B) 17H01695.
TM is supported by JST ACT-I No.JPMJPR16UP and a JSPS Grant-in-Aid
for Young Scientists (B) 17K12637.

\appendix*

\section{Proof of the Soundness Property}
First, for any $n$-qubit quantum state $\sigma$, we can show that
\begin{equation}
{\rm Tr}[ (T^{\otimes k}\otimes \Pi_G^{\bot})\sigma^{\otimes k +1}] \le 
\frac{1}{2n^2},\label{eq} 
\end{equation}
where $T$ is a POVM which corresponds to a pass by the honesty test
and 
\[ \Pi_G^\bot = I^{\otimes n} - |G\rangle\langle G|. \]
Since
\[
{\rm Tr}(T\sigma) = \frac{1}{2} + \frac{1}{2}\langle G | \sigma | G \rangle,
\]
we can say that
\begin{eqnarray*}
{\rm Tr}(\Pi_G^{\bot}\sigma ) &=& 1 - \langle G | \sigma | G \rangle\\
&=& 2(1-{\rm Tr}(T\sigma)).
\end{eqnarray*}
Thus, we have
\begin{eqnarray*}
{\rm Tr}[(T^{\otimes k}\otimes \Pi_G^\bot)\sigma^{\otimes k+1} ]
& = & {\rm Tr}(T\sigma)^k{\rm Tr}(\Pi_G^\bot \sigma)\\
&=& 2{\rm Tr}(T\sigma)^k (1 - {\rm Tr}(T\sigma)).
\end{eqnarray*}
When ${\rm Tr}(T\sigma) = \frac{k}{k+1}$,
the above takes the maximum value
\[
2\left( \frac{k}{k+1} \right)^k \left(1-\frac{k}{k+1}\right)
\le \frac{2}{k+1}\le \frac{1}{2n^2}. 
\]
The remaining $(k+1)$ qubits quantum state $\rho$
after the trace-out
can be obtained as follows by using 
the quantum de Finetti theorem with respect to the one-way LOCC norm
\begin{eqnarray*}
\lefteqn{{\rm Tr}[(T^{\otimes k}\otimes \Pi_G^{\bot})\rho ]}\\
& \le &\int d\mu (\sigma) {\rm Tr}[(T^{\otimes k}\otimes \Pi_G^{\bot})
\sigma^{\otimes k+1}] + \frac{1}{2}\sqrt{\frac{2k^2n\ln 2}{m}}\\
&\le &  \frac{1}{2n^2} + \frac{1}{2n^2} = \frac{1}{n^2}.
\end{eqnarray*}
Since 
\[ {\rm Tr}[(T^{\otimes k}\otimes \Pi_G^{\bot})\rho ]
= {\rm Tr}(\Pi_G^\bot \rho_{\rm comp}){\rm Tr}(T^{\otimes k}\rho),
\]
${\rm Tr}(\Pi_G^\bot \rho_{\rm comp})\ge \frac{1}{n}$ implies that
${\rm Tr}(T^{\otimes k}\rho)\le \frac{1}{n}$. This means that
If Alice accpets, then it holds that
$\langle G|\rho_{\rm comp} | G \rangle \ge 1 - \frac{1}{n}$
with probability $1-\frac{1}{n}$ at least. Thus the soundness
holds.

\end{document}